\documentclass{elsart}
\usepackage{epsfig,amssymb}

\newcommand{\dif}{\mathrm{d}}

\begin{document}

\begin{flushright}
HIP-2005-14/TH
\end{flushright}

\begin{frontmatter}

\title{Anisotropy of flow and the order of phase transition in relativistic
       heavy ion collisions}
\author{Pasi Huovinen}
\ead{pasi.huovinen@phys.jyu.fi}
\address{Helsinki Institute of Physics, P.O.\ Box 64,
         FIN-00014 University of Helsinki, Finland 
	 \\ and \\
	 Department of Physics, P.O.\ Box 35,
         FIN-40014 University of Jyv\"askyl\"a, Finland}

\begin{abstract}
Using a hydrodynamical model we study how the order of phase
transition in the equation of state of strongly interacting matter
affects single particle spectra, elliptic flow and higher order
anisotropies in Au+Au collisions at RHIC ($\sqrt{s_{\mathrm{NN}}}=200$
GeV energy). We find that the single particle spectra are independent
of the order of phase transition and that the fourth harmonic
$v_4(p_T)$ shows only a weak dependence in the $p_T$ region where
hydrodynamics is expected to work. The differential elliptic flow,
$v_2(p_T)$, of baryons shows the strongest dependence on equation of
state. Surprisingly the closest fit to data was obtained when the
equation of state had a strong first order phase transition and a
lattice inspired equation of state fits the data as badly as a purely
hadronic equation of state.
\end{abstract}

\end{frontmatter}

\section{Introduction}

In non-central heavy ion collisions at the Relativistic Heavy Ion
Collider (RHIC) of BNL the particle distributions exhibit quite large
anisotropies~\cite{Phenix-v2,Phobos,STAR-compiled}. The second Fourier
coefficient of the azimuthal distribution of particles, so called
elliptic flow, has been extensively studied~\cite{flowreview} since it
is sensitive to the early dense stage of the evolution~\cite{Sorge}.
Recently also higher harmonics have been
measured~\cite{STAR-compiled,STAR-v4}. It has been claimed that they
should be even more sensitive to the initial configuration of the
system~\cite{Kolbv4}.

Ideal fluid hydrodynamics has been particularly successful in
describing the observed anisotropy of particles at low $p_T$ in
minimum bias collisions~\cite{review,Peter-review}. This success has
been interpreted as a sign of formation of thermalized matter rapidly
after the primary collision~\cite{Heinz}. Studies of both single
particle spectra and anisotropies have also shown that a reasonable
reproduction of data favours an Equation of State (EoS) of strongly
interacting matter with a phase transition~\cite{Derek-RQMD,Heinz2}.

The lattice QCD calculations of the EoS of strongly interacting matter
support such a scenario by predicting a phase transition at
$T_c\approx 170$ MeV temperature. For a physical scenario of two light
and one heavier quark, the phase transition is predicted to be a
smooth crossover at small values of baryochemical potential. Contrary
to naive expectations, lattice QCD predicts that pressure and energy
density do not reach their ideal Stefan-Boltzmann values immediately
above the critical temperature, but approach them slowly~\cite{Karsch}.

At mid-rapidity at collisions at RHIC, the net baryon density is small
and the relevant EoS should exhibit a crossover transition. However,
so far all hydrodynamical calculations of elliptic
flow~\cite{Heinz,Derek-RQMD,Heinz2,letter1,letter2,alkutilat,Hirano1,Hirano,Peter-Ralf}
have used an EoS with a strong first order phase transition and ideal
parton gas to describe the plasma phase. The usual point of view has
been that it is unlikely that the details of phase transition would
lead to significant dynamical effects~\cite{Peter-review}. This
standpoint has been supported by the early
calculations~\cite{Blaizot,Dirk} where it was found that the width of
the phase transition region, $\Delta T$, had only little effect on the
final flow pattern in one dimensional flow. Thus it was considered
safe to claim that the final particle distributions would not be
sensitive to $\Delta T$ either.

However, full three dimensional expansion is more complicated than one
dimensional. It is known that in three dimensional expansion the
differential elliptic anisotropy, $v_2(p_T)$, of (anti)protons is
sensitive to the existence of phase transition and its latent
heat~\cite{Derek-RQMD,Heinz2,letter2}. The anisotropy of flow might
thus be sensitive to other details of phase transition as well. In
this paper we address this possible sensitivity. We use a
hydrodynamical model to calculate single particle spectra, elliptic
flow and higher order anisotropies in $\sqrt{s_{\mathrm{NN}}} = 200$
GeV Au+Au collisions using four different EoSs with different phase
transitions and plasma properties. As a representative of lattice QCD
results, we use an EoS based on the thermal quasiparticle model of
Schneider and Weise~\cite{Schneider} (EoS\,qp). This model is tuned to
reproduce the lattice QCD EoS and provides a method to extrapolate the
results to physical quark masses. To facilitate comparison with
earlier calculations we use as reference points the EoSs Q and H used
in Refs.~\cite{Heinz,Heinz2,letter1,letter2,alkutilat}. EoS\,Q has a
first order phase transition between hadron gas and an ideal parton
gas whereas EoS\,H is a hadron gas EoS without any phase
transition. To study the effects of the order of phase transition and
slow approach to the Stefan-Boltzmann limits separately we also use a
simple parametrisation for an EoS (EoS\,T) where the hadron gas and
ideal parton gas phases are connected using a hyperbolic tangent
function. Such an EoS has a smooth crossover transition but the plasma
properties approach their ideal values much faster than in EoS\,qp.

We find that the main sensitivity to the details of an EoS lies in the
differential elliptic flow of heavy particles ($m \gtrsim 1$ GeV)
where EoS\,Q with a first order phase transition leads to an
anisotropy closest to the data. Surprisingly, the lattice inspired
EoS\,qp reproduces the data as badly as purely hadronic EoS. EoS\,T
with a crossover transition leads to almost as good results as
EoS\,Q. Thus hydrodynamical description of elliptic flow does not
require a strong first order phase transition, but it does require
sufficiently large increase in entropy and energy densities within
sufficiently small temperature interval.

\section{Equation of State}

Until recently the lattice QCD calculations were restricted to
vanishing net baryon densities, $\mu_B = 0$. Even if there are some
recent results for $\mu_B \neq 0$~\cite{Karsch}, we limit our
discussion to zero net baryon density for the sake of simplicity.
Since we are interested in the behaviour of the collision system at
midrapidity at RHIC where net baryon density is small, this
approximation is unlikely to cause a large effect. Thermal models
suggest that around phase transition temperature, the baryon chemical
potential is below 50 MeV~\cite{PBM} corresponding to a quark chemical
potential of about 15 MeV. At these small values of $\mu$ the critical
temperature is expected to change by less than a percent from that at
$\mu = 0$~\cite{Karsch}. One of our EoSs (EoS Q, see below) also
includes extension to non-zero baryon densities. We have checked that
for this EoS, the results obtained when the finite baryon density is
included in the EoS or approximated by zero, differ by less than two
percents. Even if we do not include finite baryon density to the EoS,
we still have finite baryon current in our hydrodynamical calculation.
This allows us to have different baryon and anti-baryon yields at
freeze-out and thus finite net proton yields.

So far the lattice QCD calculations with quarks must be done using
unphysically large quark masses. The calculated equation of state must
therefore be extrapolated to physical quark mass values.  For this
purpose we use the thermal quasiparticle model of Schneider and
Weise~\cite{Schneider}. In this model the lattice QCD results are
described in terms of quasiparticles with temperature dependent
effective masses and effective number of degrees of freedom. In this
approach the EoSs obtained in lattice calculations for pure
glue~\cite{lat1} and different number of quark flavours~\cite{lat2}
are well reproduced.  Since the mass of quarks is an explicit
parameter in this model, it is easy to extrapolate the results to
physical quark masses. Here we use the quasiparticle EoS for two light
quark flavours ($m_{u,d} = 0$) and a heavier strange quark
($m_s \simeq 170$ MeV) to describe the plasma phase of an EoS\,qp.

\begin{figure}
  \begin{center}
    \epsfxsize 12cm \epsfbox{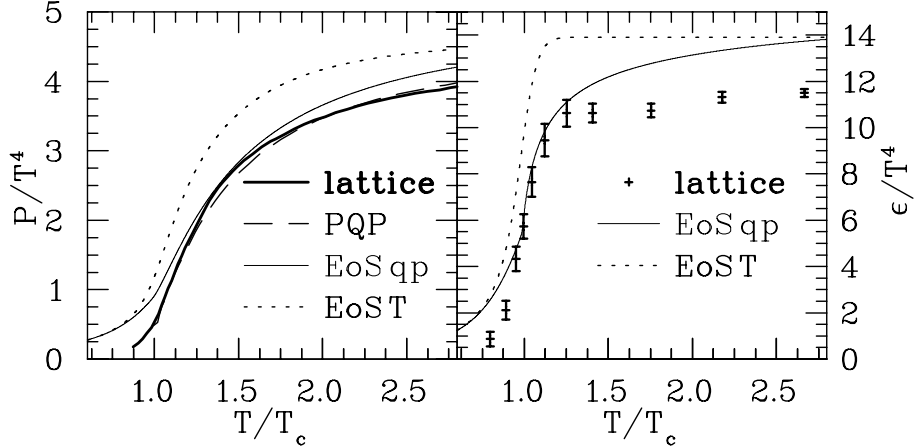}
  \end{center}
 \caption{The lattice results for pressure (left panel) and energy
          density (right panel)~\cite{lat2} compared to the quasiparticle
          model with quark masses as used in the lattice calculation (PQP),
          quasiparticle model with physical quark masses (EoS\,qp) and a
          parametrized EoS\,T (introduced later in the text). The lattice
          result for pressure is extrapolated to continuum limit by
          $p_{\mathrm{cont}}\approx 1.1 p_{\mathrm{lat}}$~\cite{Schneider}.}
  \label{hilakuva}
\end{figure}

The quasiparticle model is compared to the lattice results in
Fig.~\ref{hilakuva}, where pressure and energy density are shown as a
function of temperature. The lattice result for pressure~\cite{lat2}
is extrapolated to the continuum limit by assuming a 10\% correction,
i.e., $p_{\mathrm{cont}}\approx 1.1
p_{\mathrm{lat}}$~\cite{Schneider}, whereas the result for energy
density is shown without such an extrapolation. When the quark masses
in the quasiparticle model are set temperature dependent as in the
lattice calculations, $m_q = 0.4T$ (light quarks) and $m_s = 1.0T$
(heavy quark), the lattice pressure is nicely reproduced (PQP, dashed
line). When physical quark masses are chosen (EoS\,qp, thin solid
line), the pressure is larger than with temperature dependent masses.

There is no quasiparticle result with temperature dependent masses
available for energy density, but comparison of quasiparticle model
with physical quark masses (EoS\,qp) to the lattice shows nice
reproduction of the lattice energy density just above $T_c$ but much
larger density above $1.5T_c$. This can be partly explained by the
missing extrapolation to continuum limit of the lattice result. If one
assumes similar 10\% correction than for pressure, the difference
between lattice and EoS\,qp is quite similar for both pressure and
energy density at high temperature. The parametrized EoS\,T (see later
in the text) is included for comparison's sake and is shown to lead to
much larger pressure and energy density than lattice calculations.

The large difference between EoS\,qp and lattice below $T_c$ is
intentional and not related to the quasiparticle model.
In the present lattice simulations pions
turn out too heavy and therefore their contribution to pressure and
entropy is strongly suppressed. Thus one may expect the lattice
calculations to give too small pressure and entropy density in the
hadronic phase below $T_c$. The quasiparticle model reproduces also
this feature of the lattice EoS and one has to describe the hadronic
phase using another model.

We adopt the usual approach of using an EoS of noninteracting hadron
resonance gas to describe the hadronic phase. It has been shown that
such an EoS describes interacting hadron gas reasonably well at
temperatures around pion mass~\cite{Prakash} and that the hadron
resonance gas approach reproduces the lattice results below $T_c$ if
the same approximations are used in both~\cite{Redlich}. The
properties of hadron resonance gas depend on the number of particles
included in the model. Here we include all the strange and non-strange
particles and resonances listed in the Particle data Book up to 2 GeV
mass. The details of constructing this EoS can be found in
Ref.~\cite{Josef}.

To circumvent our ignorance of the behaviour of the EoS around $T_c$,
we use the approach outlined in Ref.~\cite{Renk}: We use the hadron
resonance gas EoS up to a temperature $T_c-\Delta T$, the
quasiparticle EoS above $T_c$ and interpolate smoothly between these
two regimes. In practice we choose the values $T_c=170$ MeV and
$\Delta T=5$ MeV and connect the entropy densities of both models
using a polynomial function. We require that the first, second and
third temperature derivatives of entropy density are continuous to
approximate a smooth crossover from hadronic to plasma phase. Below
$T_c-\Delta T$ we use the hadron resonance gas values for pressure and
energy density. Above this limit we obtain $P(T)$ and $\varepsilon(T)$
by using the thermodynamical relations $\dif P = s\,\dif T$ and
$\varepsilon = Ts-P$. This EoS is called EoS\,qp in the following.

For comparison's sake we also carry out the calculations using the
EoS\,Q and H used in
Refs.~\cite{Heinz,Heinz2,letter1,letter2,alkutilat}. EoS H is a purely
hadronic EoS without any phase transition. It is constructed by
extending the previously described hadron resonance gas EoS to
arbitrarily high temperatures. EoS\,Q, on the other hand, is inspired
by a bag model and contains a first order phase transition from hadron
gas to ideal parton gas. The hadron phase is again described by an
hadron resonance gas and the plasma phase by a gas of ideal massless
quarks and gluons with a bag constant. To approximate the effect of
the finite strange quark mass we use the number of quark flavours
$N_f=2.5$. The phase boundary is determined using the Gibbs criterion
$p_{\mathrm{HG}}(T_c) = p_{\mathrm{QGP}}(T_c)$ and the two phases are
connected using the Maxwell construction at $T_c=165$ MeV.  Details of
constructing these two EoSs can be found in Ref.~\cite{Josef}.

\begin{figure}
  \begin{center}
    \epsfxsize 12cm \epsfbox{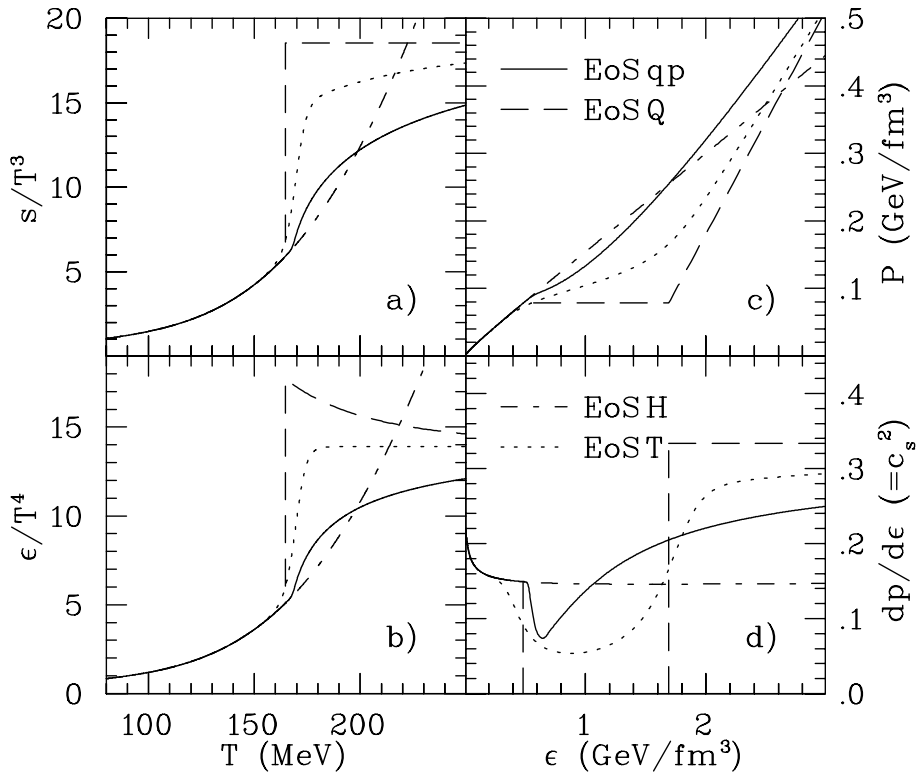}
  \end{center}
  \caption{(a) The entropy density divided by $T^3$ and (b) the energy
           density divided by $T^4$ as functions of temperature, (c)
           the pressure and (d) velocity of sound squared as functions
           of energy density in the EoSs qp (quasiparticle EoS), Q
           (ideal parton gas with first order phase transition), H
           (hadron resonance gas) and T (ansatz with crossover).}
  \label{eoskuva}
\end{figure}

As can be seen in Fig.~\ref{eoskuva} the quasiparticle and bag model
inspired EoSs (qp and Q, respectively) lead to quite different
behaviour around critical temperature. Since we want to study the
effects of the order of phase transition and slow approach to ideal
Stefan-Boltzmann values separately, we construct yet another EoS.  We
follow the idea presented in Ref.~\cite{Blaizot} and connect the
hadron and parton phases of the EoS by a hyperbolic tangent function.
We refine this approach by using hadron resonance gas EoS instead of
ideal pion gas to describe the hadron phase. In Ref.~\cite{Blaizot}
the entropy densities of hadron and parton phases are connected in
this way. This leads to similar behaviour of energy density than the
Maxwell construction of EoS\,Q --- above $T_c$ energy density rises
above the ideal Stefan-Boltzmann limit and approaches the ideal values
from above. There is no sign of this kind of behaviour in the lattice
results. Therefore we use hyperbolic tangent to connect the energy
density of the different phases instead of entropy density. Energy
density is given by
\begin{equation}
  \varepsilon(T) = \frac{1}{2}\left[\varepsilon_{\mathrm{HRG}}(T)
                   \left(1-\tanh\frac{T-T_c}{\Delta T}\right)
                  +\frac{169}{120}\pi^2T^4
                   \left(1+\tanh\frac{T-T_c}{\Delta T}\right)\right]
\end{equation}
where the latter term is the energy density of ideal parton gas with 3
colours and 2.5 quark flavours. We use again $T_c=170$ MeV and make
the crossover rapid by choosing $\Delta T = 5$ MeV. After obtaining
$\varepsilon(T)$ we again use standard thermodynamical relations,
$(\partial S/\partial E)_{N,V} = 1/T$ and $P = Ts-\varepsilon$, to
obtain entropy density and pressure as a function of temperature. This
EoS is called EoS\,T in the following.

All four EoSs are compared in Fig.~\ref{eoskuva} where entropy and
energy density are shown as functions of temperature and pressure and
the square of the speed of sound are shown as functions of energy
density for each EoS. As can be seen the behaviour of the lattice
inspired EoS\,qp is quite different from the previously used EoS\,Q
with a first order phase transition. The latter has a relatively large
latent heat of 1.15 GeV/fm$^3$ whereas in the former the region where
the speed of sound is small and the EoS soft is much smaller. The
parametrised EoS\,T is a compromise between these two. It can also be
seen that above the phase transition region the EoS\,Q has the largest
speed of sound and is therefore hardest whereas the EoS\,H without
phase transition is softest.

It is worth noticing that EoS\,qp depicts a smaller rise in both
energy and entropy densities around $T_c$ than what could be expected
from lattice calculations. This is not a property of the quasiparticle
model used here, but due to the use of hadron resonance gas EoS below
$T_c$. As mentioned before, lattice calculations lead to too high pion
mass and correspondingly too small densities below $T_c$. If realistic
pion masses are used in hadron resonance gas, its pressure and
densities are well above lattice results below $T_c$.

\section{Initialization}
  \label{init}

We use the same boost-invariant hydrodynamic code than in
Refs.~\cite{letter1,letter2,alkutilat} and described in detail in
Ref.~\cite{Peterinpitka}. To fix the parameters of the model, we
require that the model reproduces the $p_T$ spectra of pions and net
protons ($p - \bar{p}$) in most central collisions and the centrality
dependence of pion multiplicity at midrapidity. We use net protons
instead of protons and/or antiprotons because our model assumes
chemical equilibrium to hold down to kinetic freeze-out temperature
and is unable to reproduce proton and antiproton yields
simultaneously.

Some parametrisations to fix the initial density distributions were
explored in Ref.~\cite{alkutilat}. None of them reproduces the
observed centrality dependence of multiplicity, but a linear
combination of them does. Here we use the same combination than in
Refs.~\cite{Heinz,Heinz2}. The local entropy density is taken to scale
with a linear combination of the density of participants and binary
collisions in the transverse plane with weights of 0.75 and 0.25,
respectively. This kind of scaling can be interpreted as particle
production from ``soft'' and ``hard'' processes.  For the sake of
simplicity, the initial baryon number density is taken to scale with
the number of participants. The initial time of the calculation,
$\tau_0 = 0.6$ fm/$c$, is taken from earlier calculations for
$\sqrt{s_{\mathrm{NN}}}=130$ GeV energy~\cite{letter1,letter2}.

The freeze-out energy density is chosen to reproduce the slopes of
pion and net proton spectra in most central collisions (see upper left
panel of Fig.~\ref{spektrit}). The stiffer the EoS, the sooner,
i.e.\ at higher density, the necessary flow velocity to fit the
spectra has been built up. We find that effectively the EoS\,qp is
stiffest since it requires the highest decoupling energy density 
$\varepsilon_{fo} = 0.14$ GeV/fm$^3$ ($\langle T_{fo}\rangle = 141$
MeV) to fit the data. Even if the ideal parton gas EoS is stiff, the
mixed phase of the EoS\,Q makes it effectively the softest EoS
here. The stiffening of the phase transition region and softening
of the plasma phase in the EoS\,T cancel each other. It is almost
as soft as EoS\,Q and can use the same freeze-out energy density
$\varepsilon_{fo} = 0.08$ GeV/fm$^3$ ($\langle T_{fo}\rangle = 130$
MeV). The purely hadronic EoS H is in between these extremes and
requires $\varepsilon_{fo} = 0.10$ GeV/fm$^3$ ($\langle T_{fo}\rangle
= 135$ MeV) to fit the data. As can be seen in the upper left panel of
Fig.~\ref{spektrit}, these choices of $\varepsilon_{fo}$ allow all
EoSs to fit the data equally well.

\section{Results}

     \subsection{$p_T$-spectra}

\begin{figure}
 \begin{minipage}{7cm}
   \begin{center}
     \epsfxsize 65mm \epsfbox{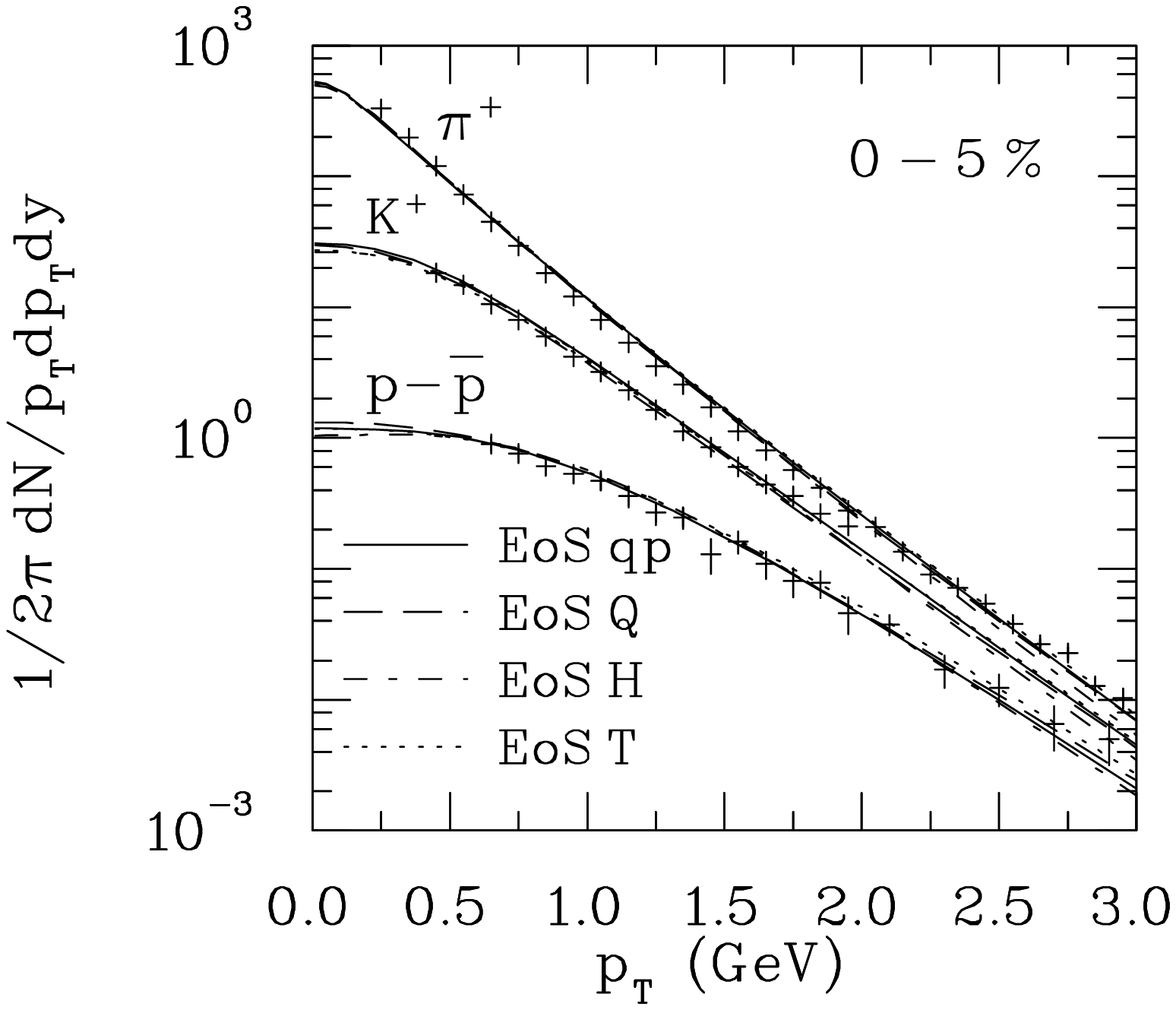}
   \end{center}
 \end{minipage}
   \hfill
 \begin{minipage}{7cm}
   \begin{center}
     \epsfxsize 65mm \epsfbox{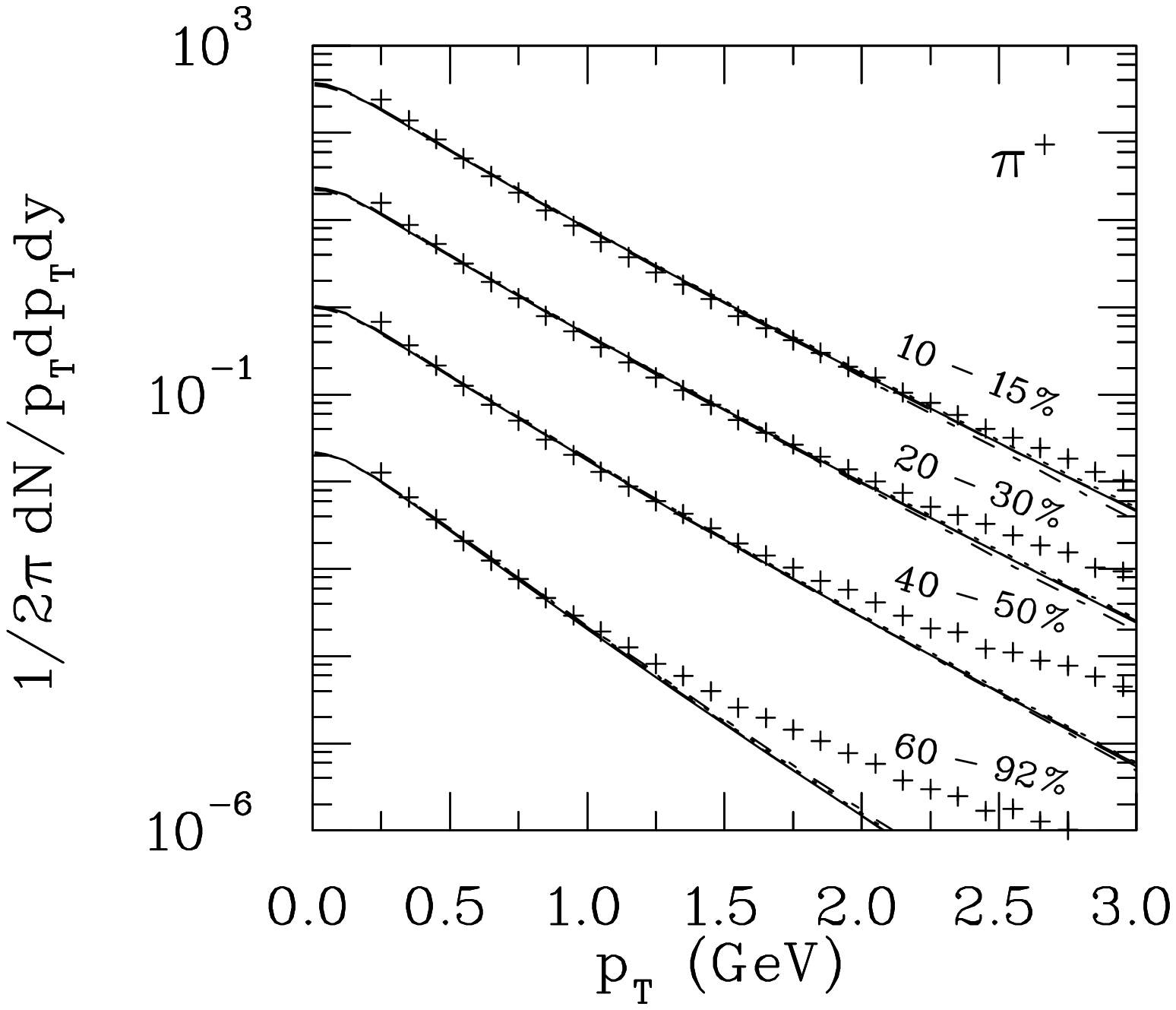}
   \end{center}
 \end{minipage}
   \hfill
 \begin{minipage}{7cm}
   \begin{center}
     \epsfxsize 65mm \epsfbox{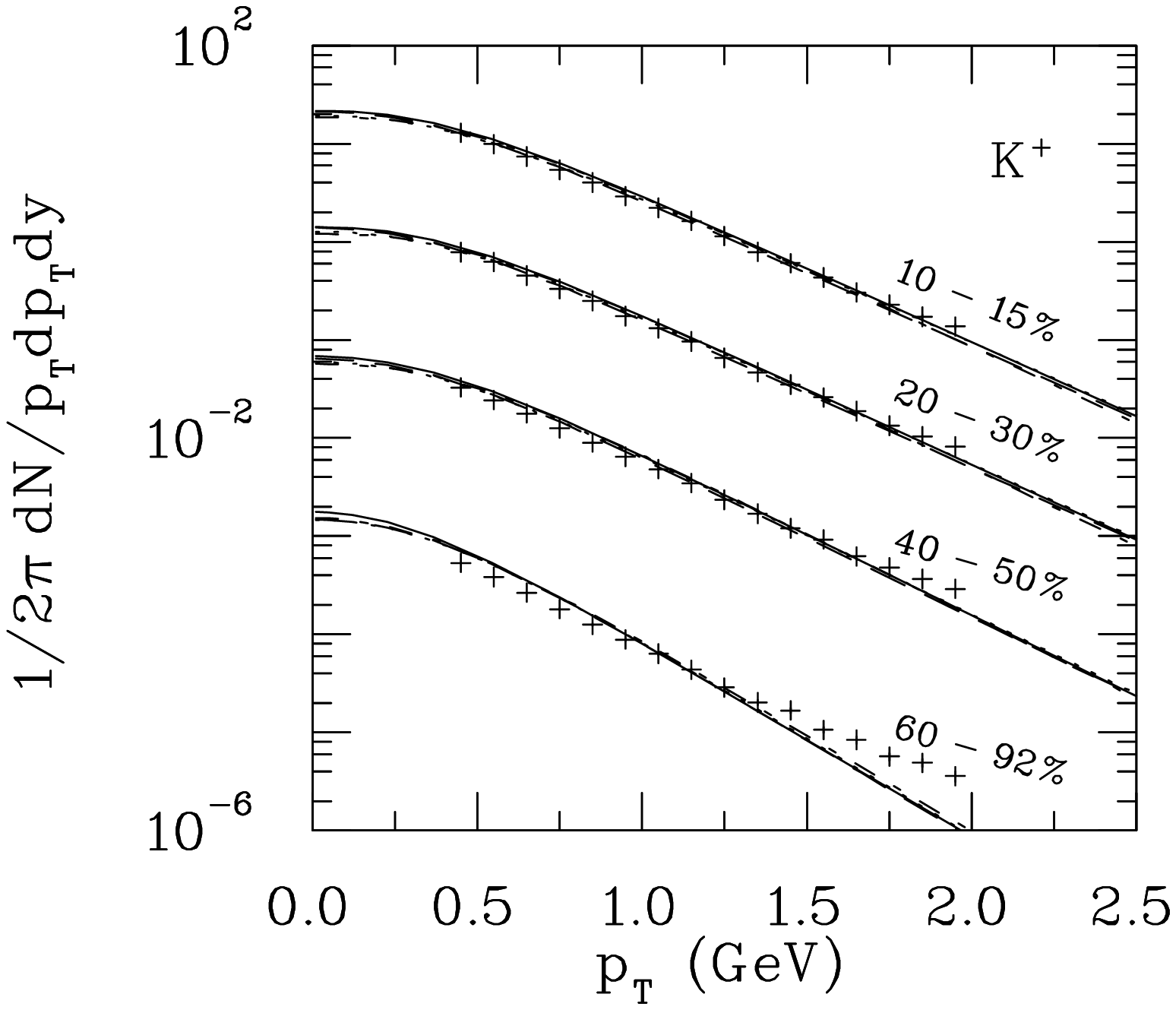}
   \end{center}
 \end{minipage}
   \hfill
 \begin{minipage}{7cm}
   \begin{center}
     \epsfxsize 65mm \epsfbox{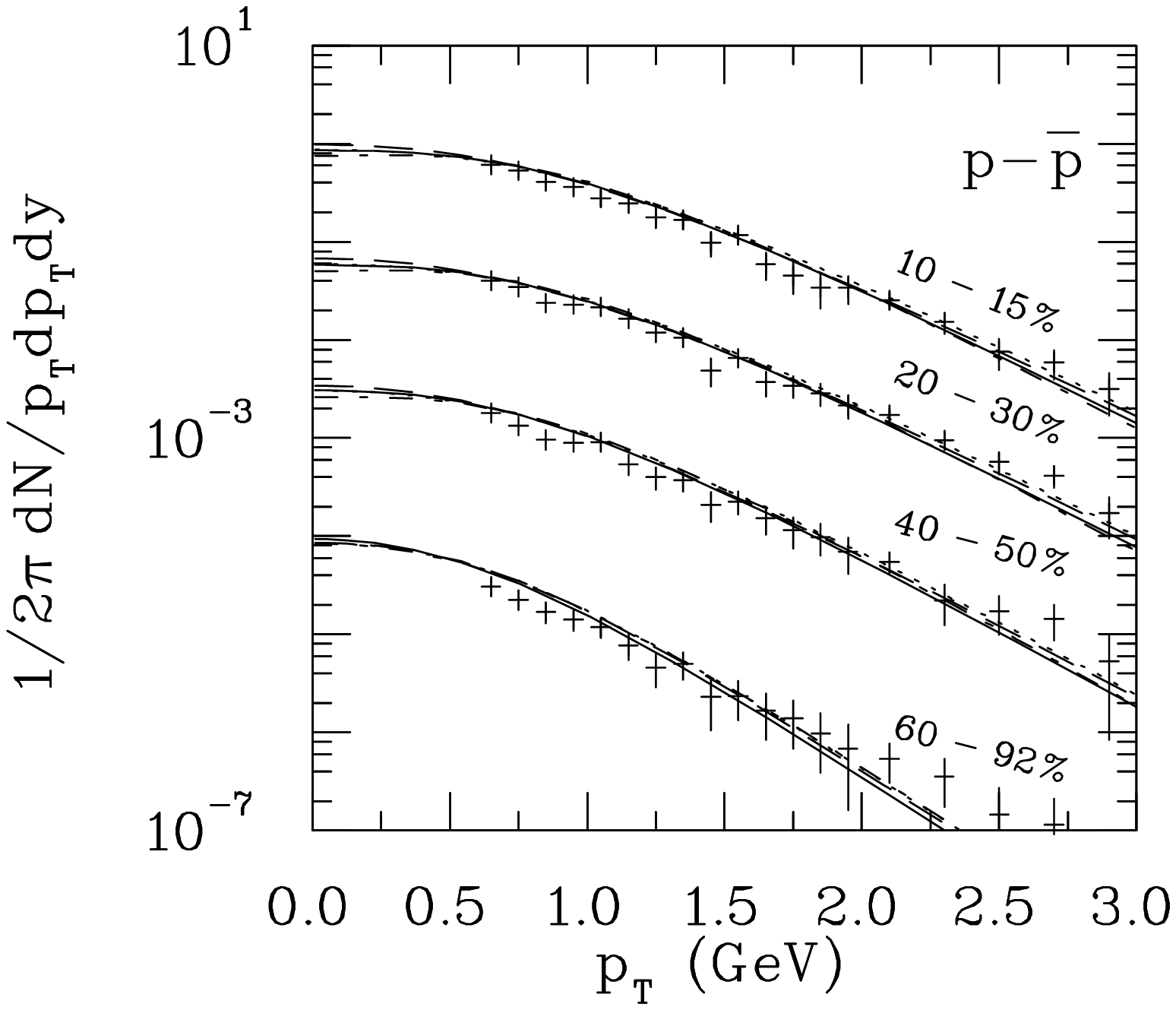}
   \end{center}
 \end{minipage}
 \caption{Pion ($\pi^+$), kaon ($K^+$) and net proton ($p-\bar{p}$)
          $p_T$-spectra in most central (top left) and semi-central to
          peripheral Au+Au collisions at $\sqrt{s_{AA}} = 200$ GeV
          compared with hydrodynamical calculations using four
          different EoSs. The data was taken by the PHENIX
          collaboration~\cite{Phenix-pt}. For clarity the spectra at
          centralities 20 - 30\%, 40 - 60\% and 60 - 80\% are scaled
          by factors 10$^{-1}$, 10$^{-2}$ and 10$^{-3}$, respectively.}
 \label{spektrit}
\end{figure}

The transverse momentum spectra for pions, kaons and net protons for
various centralities are shown in Fig.~\ref{spektrit}. Pion and net
proton spectra in most central collisions were used to fix the
parameters of the model, but the data are well reproduced at other
centralities too. Only at the most peripheral collisions the data tend
to favour flatter spectra than calculated.

The kaon spectra was not taken into account when choosing the
parameters and the calculated spectra is a prediction in most central
collisions too. The fit to data is surprisingly good when one takes
into account that the freeze-out temperature is well below the
$T_{\mathrm{chem}}\approx 174$ MeV chemical freeze-out temperature
where particle yields are fixed~\cite{PBM}.

The net proton spectra is well reproduced up to $p_T = 2.5$ - 3.0 GeV
except in the most peripheral collisions, where the data begins to
deviate form the calculation at lower $p_T$. It is worth noticing that
we are able to fit the $p_T$ spectra of net protons without any
initial transverse velocity field whereas the fit of protons in
Ref.~\cite{Peter-Ralf} required a non-zero initial transverse
velocity. One reason for this is that due to larger errors, it is
easier to fit the net-proton than proton spectra. The main cause is,
however, the different EoS in the hadronic phase. In~\cite{Peter-Ralf},
the authors assumed a separate thermal and kinetic freeze-outs and
only a partial chemical equilibrium in the hadronic phase whereas in
this work a full chemical equilibrium is assumed. Although the
relation between pressure and energy density is almost independent of
these assumptions, the relation between temperature and energy density
depends strongly on them~\cite{Hirano,Derek-eos}. Thus the relation
between collective and thermal motion in a hydrodynamical model
depends on the assumption of chemical equilibrium or non-equilibrium
and very different initial states can be required to fit the data.

  \subsection{Elliptic anisotropy}

\begin{figure}
  \begin{center}
    \epsfxsize 12cm \epsfbox{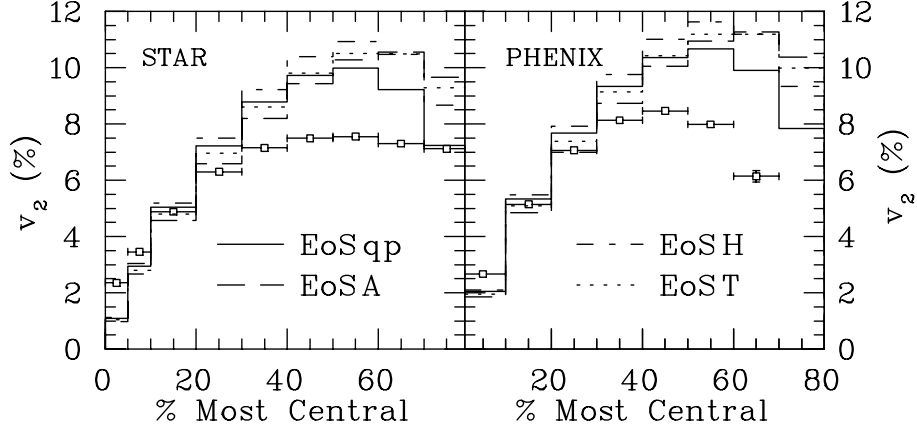}
  \end{center}
 \caption{Centrality dependence of elliptic flow of charged hadrons
          calculated using three different equations of state and
          compared with data by the STAR~\cite{STAR-v4} and the
          PHENIX~\cite{Phenix-v2} collaborations. The STAR data is
	  for $|\eta| < 1.2$ and $p_T > 0.15$ GeV and the PHENIX data
	  is for $|\eta| < 0.35$ and $0.2 < p_T < 10$ GeV.
          The same cuts have been applied to the hydrodynamic calculations.}
  \label{v2cent}
\end{figure}

The second Fourier coefficient, $v_2$, of the azimuthal distribution
of charged particles as function of centrality is shown as a histogram
in Fig.~\ref{v2cent}. Note that the data measured by the
STAR~\cite{STAR-v4} and PHENIX~\cite{Phenix-v2} collaborations have
different pseudorapidity and $p_T$ cuts. After these cuts have been
applied to the calculations, the results differ slightly. Therefore
the comparison with the data is done in two separate panels. The
agreement with data is similar to that seen in $\sqrt{s_{\mathrm{NN}}}
= 130$ GeV collisions~\cite{Derek-RQMD,letter1,QM02}: at most central
collisions ($< 10$\% of cross section, $b \lesssim 4.6$ fm) the
observed anisotropy is above the hydrodynamical result. At semicentral
collisions the calculations fit the data (10 - 30\% of cross section,
$4.6 \lesssim b \lesssim 8$ fm, depending on the EoS) and at
peripheral collisions the calculated anisotropy is well above the
observed. One possible explanation for larger observed than calculated
anisotropy in most central collisions is fluctuations in the initial
state geometry~\cite{Miller}.  The present experimental procedure
cannot distinguish between the enhancing and suppressing effects of
fluctuations on anisotropy in most central collisions and consequently
leads to too large value of $v_2$.

The sensitivity of the anisotropy to the EoS depends on centrality. In
the most central and semi-peripheral collisions EoS\,Q leads to the
lowest anisotropy and EoS\,H to the largest, but in most peripheral
collisions the lowest anisotropy is achieved using EoS\,qp. The
stiffest EoS does not always lead to the largest anisotropy and the
softest to smallest because of the interplay of collective and thermal
motion. Stiff EoS may necessitate decoupling at higher temperature
when larger thermal motion dilutes the flow anisotropy.

\begin{figure}
  \begin{center}
    \epsfxsize 12cm \epsfbox{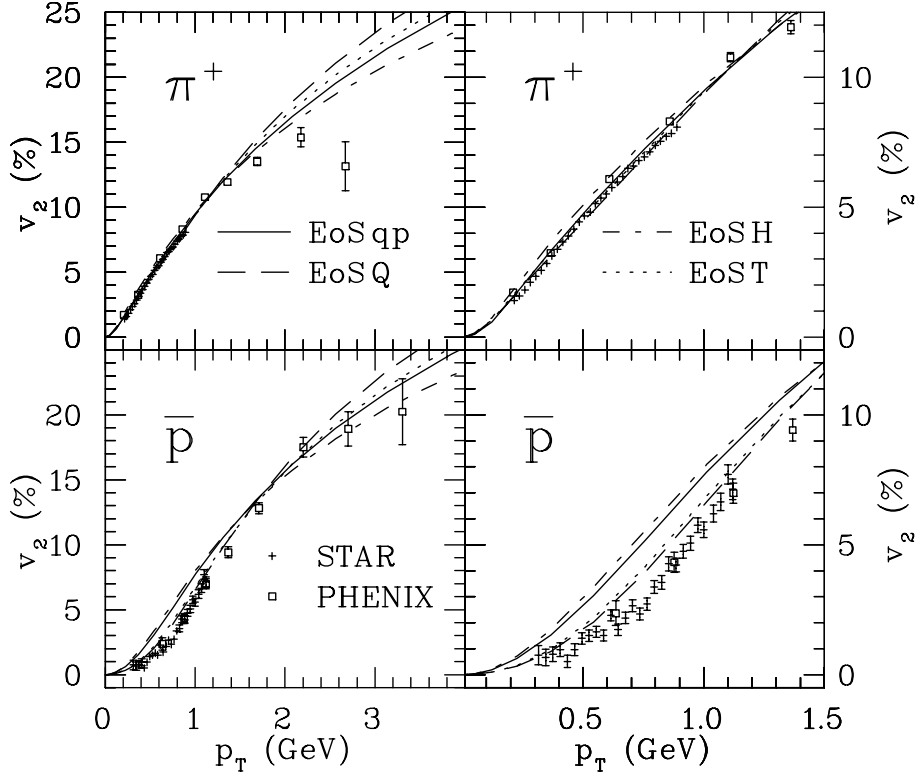}
  \end{center}
  \caption{Elliptic flow of pions and anti protons vs. transverse
          momentum in minimum bias Au+Au collisions at
          $\sqrt{s_{\mathrm{NN}}}=200$ GeV calculated using four
          different EoSs and compared with the data by the
          STAR~\cite{STAR-compiled} and PHENIX~\cite{Phenix-v2}
          collaborations. Feed-down from weak decays of strange
          baryons is included in the calculations.}
  \label{v2piap}
\end{figure}

\begin{figure}
  \begin{center}
    \epsfxsize 12cm \epsfbox{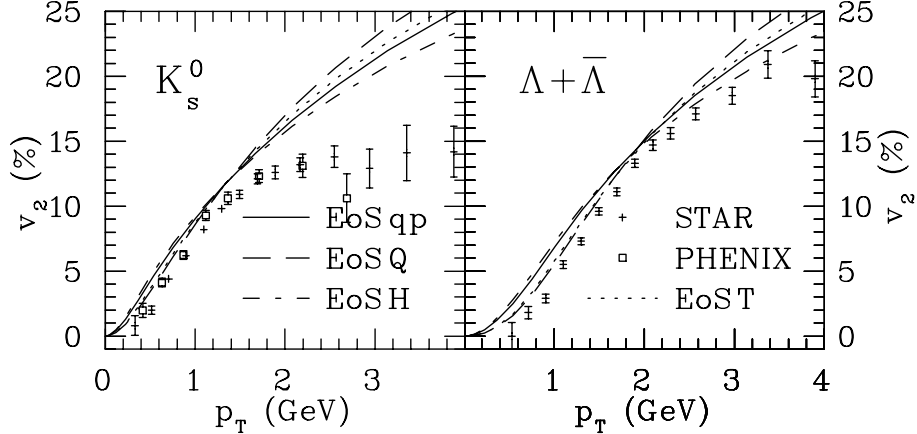}
  \end{center}
  \caption{Elliptic flow of kaons and lambdas vs. transverse momentum
          in minimum bias Au+Au collisions at
          $\sqrt{s_{\mathrm{NN}}}=200$ GeV calculated using four
          different EoSs and compared with the data by the
          STAR~\cite{STAR-compiled} and PHENIX~\cite{Phenix-v2}
          collaborations.}
  \label{v2KL}
\end{figure}

The momentum dependence of elliptic flow, $v_2(p_T)$, in minimum bias
collisions is shown in Fig.~\ref{v2piap} for positive pions and
antiprotons and in Fig.~\ref{v2KL} for neutral kaons and a sum of
lambdas and antilambdas. For pions the behaviour is similar to the
charged particle $v_2(p_T)$ at $\sqrt{s_{\mathrm{NN}}} = 130$ GeV
collisions~\cite{Heinz2,letter1}. Regardless of the EoS the calculated
anisotropy reproduces the data up to $p_T \approx 1.5$ GeV where the
data begins to saturate but the hydrodynamical curve keeps
increasing. Major differences between different EoSs are at the high
$p_T$ region where no EoS fits the data. Closer look at low $p_T$
region ($p_T < 1$ GeV) reveals that EoS\,H leads to slightly larger
$v_2$ than the other EoSs, but the difference is equal to the
difference between the STAR and PHENIX data.

The antiprotons show much stronger sensitivity to the EoS than
pions. Below $p_T = 2$ GeV the results form two groups. EoSs\,qp and H
lead to almost identical $v_2(p_T)$ which is clearly above the data
whereas EoSs\,Q and T lead to anisotropy very close to the data. The
phase transition crossover in EoS\,T is very rapid with $\Delta T = 5$
MeV. We have tested that increase in $\Delta T$ leads to larger
antiproton $v_2(p_T)$ at low $p_T$ and worse fit with the data. For
example $\Delta T = 17$ MeV moves the $v_2(p_T)$ curve roughly halfway
between results for EoS\,Q and qp. At high values of $p_T$ the order
of results is changed with EoS\,Q leading to highest and EoS\,H to the
lowest anisotropy. The antiproton data follows the hydrodynamical
calculation to much higher values of $p_T$ than the pion data. Even
the highest data point at $p_T=3.2$ GeV is fitted while using EoS\,qp
or H.

Even when EoS\,Q is used, we can not reproduce the antiproton
$v_2(p_T)$ as well as in earlier studies~\cite{letter2}. The main
reason is that in Ref.~\cite{letter2} freeze-out temperature was lower
$T_f \approx 120$ MeV, but after constraining the freeze-out to fit
the $p_T$ spectra we are forced to use higher freeze-out temperature
which does not allow as good description of the $v_2$ data.

The general behaviour of antiproton $v_2(p_T)$ suggests that the
larger the latent heat, the smaller the $v_2(p_T)$ at low $p_T$.
However, this is not the case. To test this hypothesis we used also an
EoS with a first order phase transition and larger latent heat than
EoS\,Q (2 GeV/fm$^3$ instead of 1.15 GeV/fm$^3$). The fit to
antiproton anisotropy was no better than for EoS\,Q (similar behaviour
was already seen in Ref.~\cite{Derek-RQMD} for EoSs with latent heats
0.8 and 1.6 GeV/fm$^3$).

Comparison with the strange particle data ($K_s^0$, $\Lambda +
\bar{\Lambda}$) in Fig.~\ref{v2KL} shows similar trends. The larger
the particle mass the larger the differences between EoSs at low
$p_T$. The data deviates from the overall behaviour of hydrodynamical
calculation at lower $p_T$ for mesons than for baryons -- the kaon
data deviates already around $p_T \approx 1.2$ GeV whereas
hydrodynamical calculation is close to lambda data up to $p_T \approx
3.5$ GeV. On the other hand the overall fit to data is worse for
strange than non-strange particles. Even EoS\,Q leads to calculated
anisotropy which is above the data also at low $p_T$. Smaller $v_2$ at
low $p_T$ can not be interpreted as a sign of strange particles
freezing out earlier at higher temperature. For kaons and lambdas that
would mean \emph{larger} $v_2$ at small $p_T$. The good fit to kaon
$p_T$ spectra is also against a higher kinetic freeze-out temperature
for strange particles.

As shown in Ref.~\cite{alkutilat}, the different parametrisations of
the initial state can lead to similar pion $v_2(p_T)$ but different
proton $v_2(p_T)$ in minimum bias collisions. We have checked if it
would be possible to bring the antiproton $v_2(p_T)$ down to fit the
data using EoS\,H but different initial state as speculated in
Ref.~\cite{review}. To do this we assumed that at each value of impact
parameter $b$, the initial energy density was proportional to the
density of binary collisions in the transverse plane (parametrisation
eBC of Ref.~\cite{alkutilat}), but the proportionality constant
depended on impact parameter to reproduce the observed centrality
dependence of multiplicity. Because this parametrisation led to
steeper initial gradients than our usual parametrisation, we had to
use freeze-out energy density $\varepsilon_f = 0.12$ GeV/fm$^3$
instead of $\varepsilon_f = 0.1$ GeV/fm$^3$ ($\langle T_f\rangle =
138$ MeV and $\langle T_f\rangle = 135$ MeV, respectively) to
reproduce the $p_T$ distributions of pions and net-protons. As a
result, the earlier decoupling negated the change due to the different
initial shape and the final proton $v_2(p_T)$ was almost similar to
that shown in Fig.~\ref{v2piap} and well above the data. We conclude
that the anisotropies shown in Figs.~\ref{v2piap} and \ref{v2KL} are
typical for each EoS and robust against small variations in the
initial parametrisation of the system.

  \subsection{Higher harmonics}

\begin{figure}
  \begin{minipage}{7cm}
    \begin{center}
      \epsfysize 55mm \epsfbox{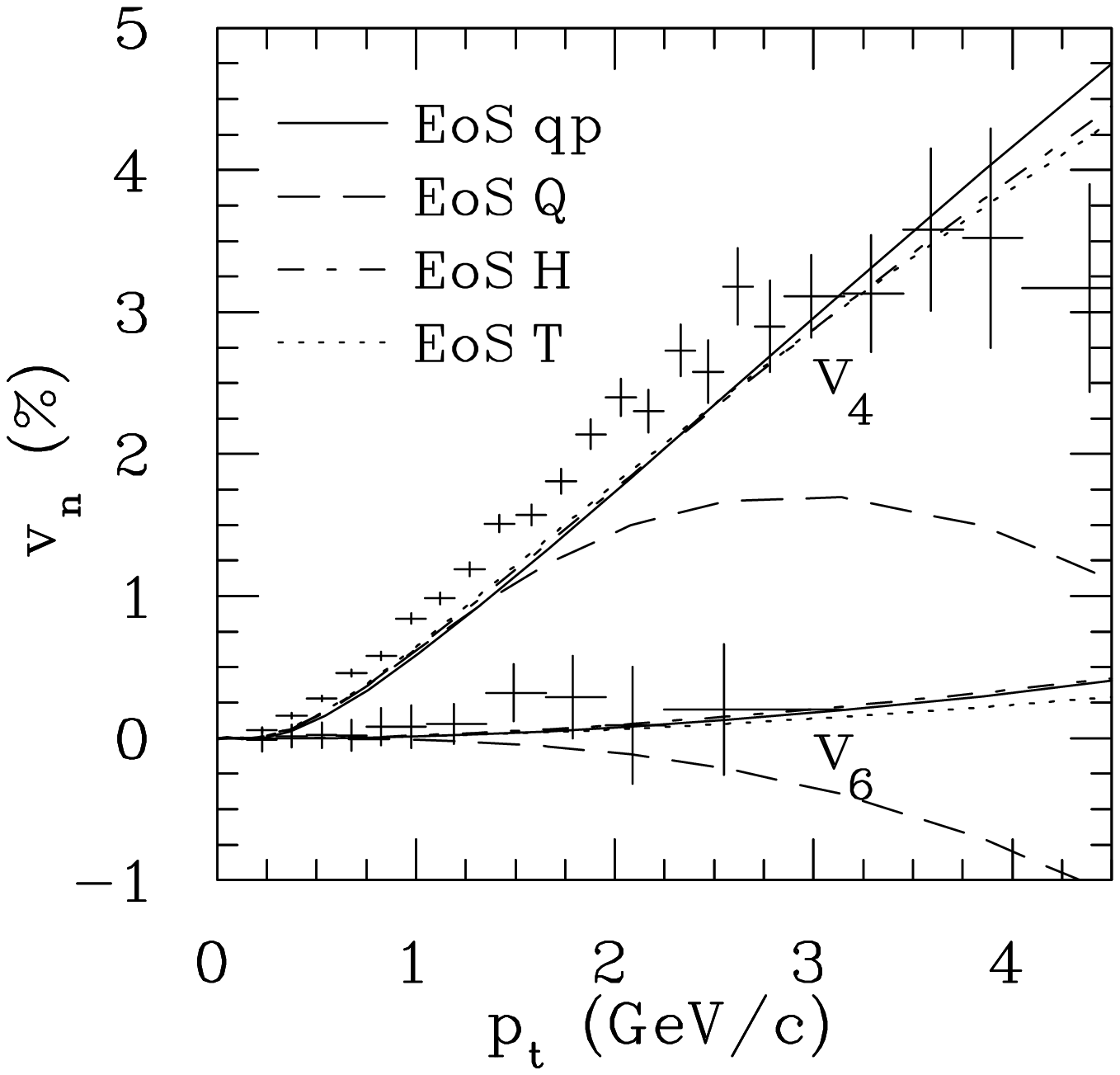}
    \end{center}
  \end{minipage}
    \hfill
  \begin{minipage}{7cm}
    \begin{center}
      \epsfysize 55 mm \epsfbox{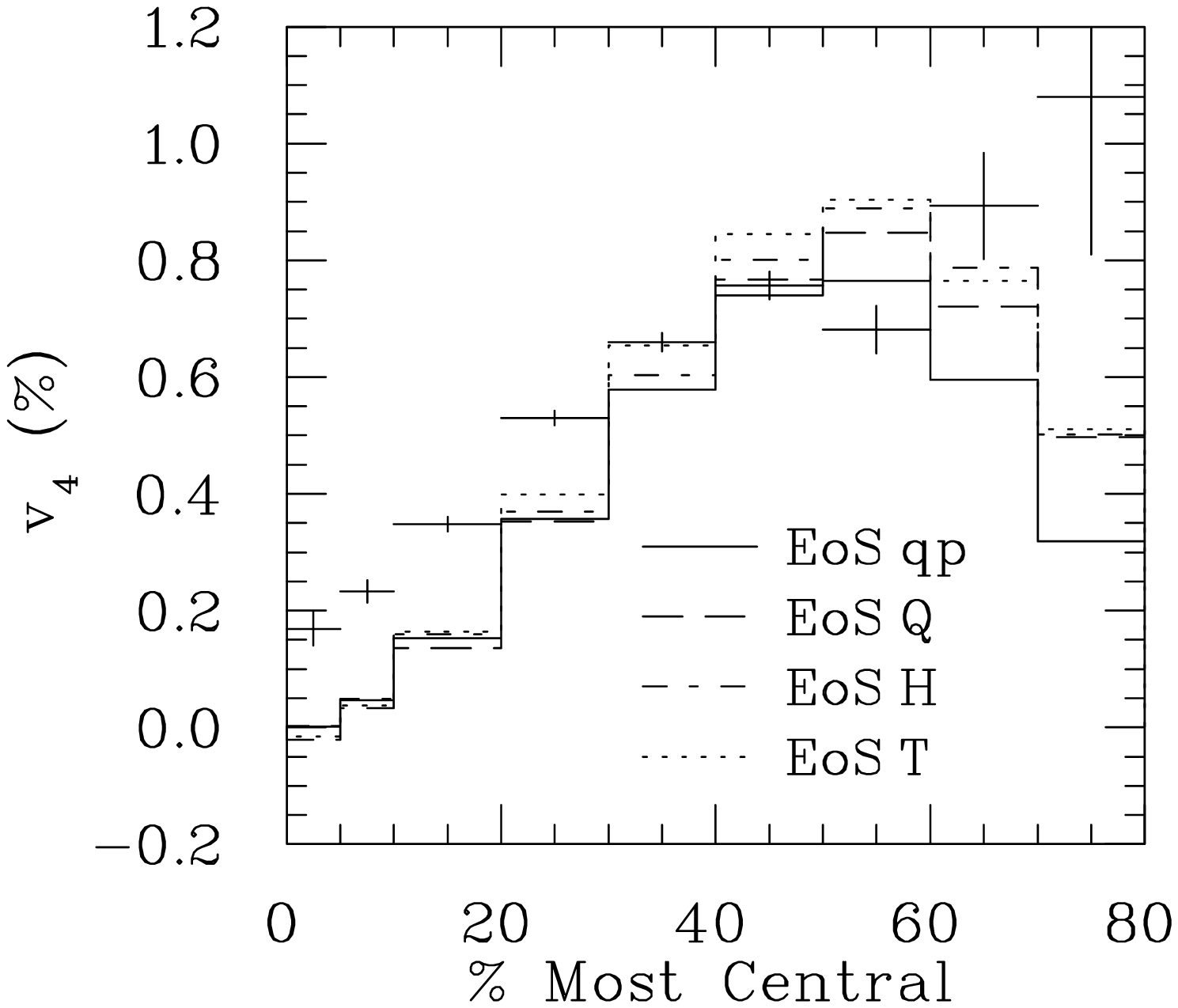}
    \end{center}
  \end{minipage}
  \caption{The fourth and sixth harmonics, $v_4(p_T)$ and $v_6(p_T)$,
           in minimum bias collisions (left) and $p_T$-averaged fourth
           harmonic, $v_4$, of the azimuthal distribution of charged
           hadrons as function of centrality (right) calculated using
           four different EoSs and compared with the STAR
           data~\cite{STAR-v4}.}
  \label{v46charged}
\end{figure}

Recently there has been interest in measuring the higher harmonics of
the azimuthal distribution of particles~\cite{STAR-compiled,STAR-v4}.
It has been proposed that these higher coefficients should be even
more sensitive to the initial configuration of the system than the
elliptic flow coefficient $v_2$~\cite{Kolbv4}. A detailed study of
these coefficients would require checking how different initial
configurations would affect these coefficients. Instead we calculate
the fourth and sixth harmonics of distribution, $v_4$ and $v_6$, using
the initial state defined above as a first attempt to see how an EoS
affects higher harmonics.

The fourth and sixth harmonics of the charged particle distribution in
minimum bias collisions, $v_4(p_T)$ and $v_6(p_T)$ as function of
transverse momenta are shown in the left panel of
Fig.~\ref{v46charged}. The EoS has significant effect only above
$p_T\approx 2$ GeV, i.e.\ in the region where $v_2(p_T)$ is no longer
reproduced by hydrodynamics. EoS\,Q leads to $v_4$ peaking around
$p_T\approx 3$ GeV whereas all the other EoSs lead to monotonous
increase of $v_4(p_T)$ with increasing $p_T$. The data, on the other
hand, increases up to $p_T\approx 3$ GeV and saturates. Except for the
high $p_T$ region all the EoSs lead to $v_4$ which is smaller than the
experimentally measured values. The measured values of the sixth
harmonics of the distribution, $v_6$, are consistent with zero,
although the errors are large enough not to exclude any of the
calculations here. The calculated values of $v_6$ are also small but
show a qualitative dependence on the EoS: EoS\,Q leads to negative
$v_6$ whereas all the other EoSs lead to positive values of $v_6$.

The centrality dependence of the $p_T$ averaged fourth harmonic $v_4$
is shown in the right panel of Fig.~\ref{v46charged}. It shows
qualitatively similar behaviour to $v_4(p_T)$. The EoS has only a weak
effect on $v_4$ except in peripheral collisions. In central and
semicentral collisions the calculated values are below the observed
ones.

\begin{figure}
  \begin{center}
    \epsfxsize 11cm \epsfbox{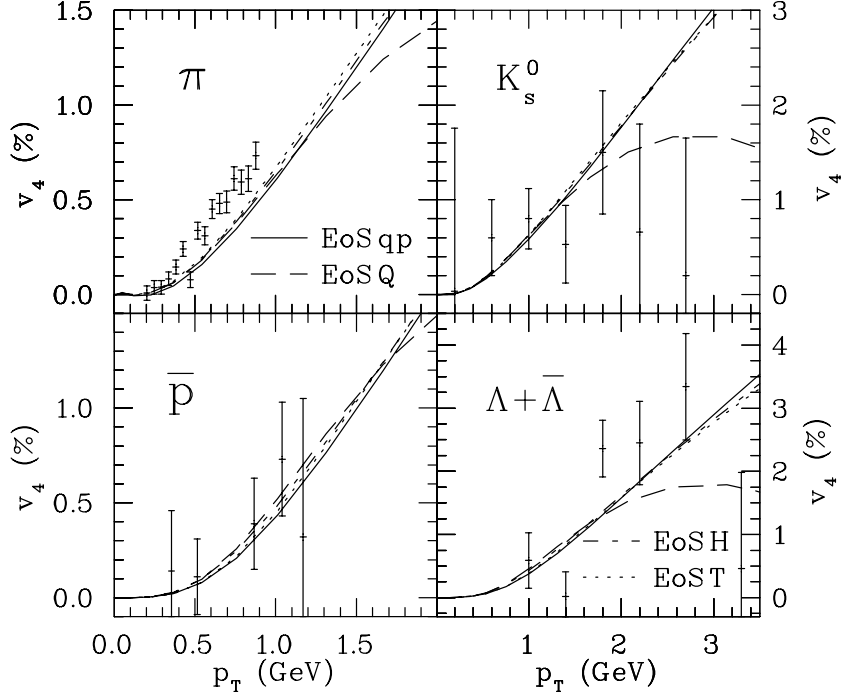}
  \end{center}
  \caption{The fourth harmonic, $v_4(p_T)$, of the azimuthal particle
           distribution of pions, anti-protons, $K_S^0$ and
           $\Lambda+\bar{\Lambda}$ for minimum bias Au+Au collisions
           calculated using four different EoSs and compared with the
           preliminary STAR data~\cite{STAR-compiled}.}
  \label{v4particles}
\end{figure}

In Fig.~\ref{v4particles} the $p_T$ dependence of fourth harmonic
$v_4$ in minimum bias collisions is shown for identified pions, kaons
($K_s^0$), antiprotons and lambdas. As was the case for charged
hadrons, the EoS has only a weak effect on results below $p_T\approx
2$ GeV. The pion data is above the hydrodynamical calculations. The
errors for other particles are large and the calculations fit the data
except at the highest $p_T$ where kaon data seems to favour EoS\,Q and
lambda data all the other EoSs.

\section{Flow on decoupling surface}

To understand how different EoSs lead to different anisotropies, we
study the properties of the freeze-out surface in Au+Au collision with
impact parameter $b=6$ fm. We try to find a set of parameters to
describe the surface similar to those presented in Ref.~\cite{STAR130}
for a blast wave model. Freeze-out temperature, average transverse
flow velocity and two anisotropy coefficients are shown in
Table~\ref{omituisuudet}. To characterise the spatial anisotropy of
the surface we generalise the usual spatial anisotropy
$\epsilon_x$~\cite{Peterinpitka} for hypersurfaces:
\begin{equation}
   \epsilon_x = \frac{\int \partial\sigma_\mu s^\mu\, (y^2 - x^2)}
                     {\int \partial\sigma_\mu s^\mu\, (y^2 + x^2)},
\end{equation}
where the usual integral over $\dif x\,\dif y$ is replaced by an
integral over space-time hypersurface and instead of energy density,
entropy density is used as a weight. To characterise the azimuthal
modulation of the flow field, we first calculate average flow velocity
as function of flow angle, $\langle v_r(\phi_v)\rangle$, where
$\phi_v = \arctan(v_y/v_x)$. We use the second Fourier coefficient
of this distribution as a measure of anisotropy of the flow field:
\begin{equation}
  a_2 = \frac{\int \dif\phi\, \langle v_r (\phi)\rangle\, \cos(2\phi)}
             {\int \dif\phi\, \langle v_r (\phi)\rangle}.
\end{equation}
This allows us to separate the spatial anisotropy from the flow anisotropy.

\begin{table}
 \begin{center}
  \begin{tabular}{|l|c|c|c|c|} \hline
                                & EoS\,qp & EoS\,Q & EoS\,H & EoS\,T \\ \hline
  $\langle T_{fo}\rangle$ (MeV) & 141     & 130    & 134    & 130    \\ \hline
  $\langle v_r \rangle$         & 0.47    & 0.47   & 0.49   & 0.49   \\ \hline
  $\epsilon_x$                  & 0.058   & 0.033  & 0.056  & 0.034  \\ \hline
  $a_2$                         & 0.027   & 0.027  & 0.025  & 0.026  \\ \hline
  \end{tabular}
  \caption{Freeze-out temperature, average transverse flow velocity,
           spatial eccentricity and flow anisotropy on the decoupling
           surface in Au+Au collision with impact parameter $b=6$ fm
           using four different EoSs.}
  \label{omituisuudet}
 \end{center}
\end{table}
      
The average flow velocity and anisotropy of the velocity field are
surprisingly similar in all four cases. The main differences at
freeze-out are freeze-out temperature and the shape of the surface.
As seen in Ref.~\cite{Lisa} where the anisotropies are studied using a
parametrisation of the freeze-out surface, at this temperature and
velocity range the lower temperature should lead to larger
anisotropies for both pions and protons. As can be expected, in
parametrisation smaller spatial anisotropy is seen to lead to smaller
$v_2$ of particles. This behaviour is different from what we see here
where EoS\,Q leads to lowest $v_2(p_T)$ at low $p_T$. Smaller spatial
anisotropy can not explain this alone, since its effect should be
cancelled by lower temperature. Also the differences between EoS\,Q
and T are such that one would expect EoS\,T to lead to lower $v_2$ for
both pions and protons, but that is not the case.

\begin{figure}
  \begin{center}
    \epsfxsize 12cm \epsfbox{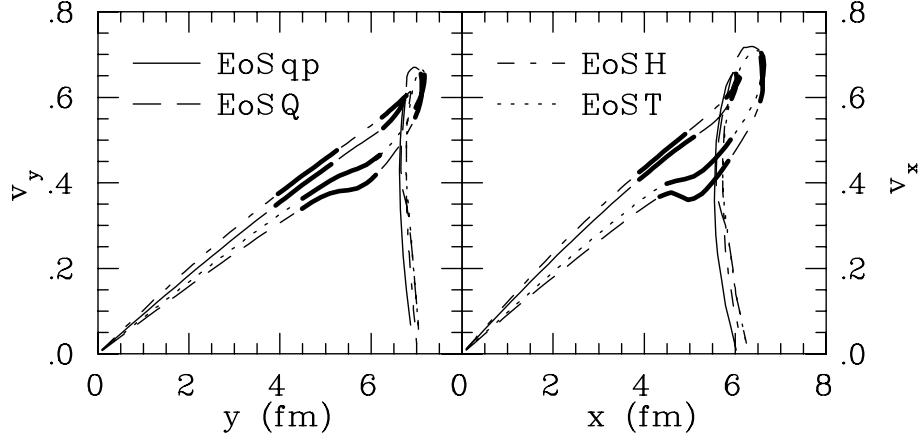}
  \end{center}
  \caption{The transverse flow velocity on decoupling surface of a
           Au+Au collision with impact parameter $b=6$ fm using four
           different EoSs. The left panel shows the velocity as
           function of $y$-coordinate when $x=0$ and the right panel
           as function of $x$ when $y=0$. The curves are divided into
           segments of thin and thick lines where each segment
           corresponds to 20\% of total entropy flowing through
           freeze-out surface.}
  \label{profiles}
\end{figure}

Clearly the average values do not characterise the flow well
enough. The reason for different anisotropies must lie in the details
of the flow profiles. To have a closer look at the properties of flow
on the decoupling surface, we have plotted the flow velocity on
decoupling surface as a function of radial coordinate in
Fig.~\ref{profiles}. In the left panel the flow velocity is shown as
function of $y$ when $x=0$ and in the right panel as function of $x$
when $y=0$.

As expected from very similar spectra and differential anisotropies,
the velocity distribution for EoSs qp and H is also close to each
other.  EoS\,Q, on the other hand, leads to different flow profile
with slower increase of velocity with increasing radius, a distinctive
``shoulder'' at $r\approx 5$ fm where the velocity can even slightly
decrease with increasing $r$ (in $x$-direction) and very rapid rise of
flow velocity close to maximum radius of the system. EoS\,T on the
other hand is somewhere in between these two with the slow rise at low
$r$ and very rapid rise at large $r$ but with much weaker structure
around $r\approx 5$ fm.

\begin{figure}
  \begin{center}
    \epsfxsize 6cm \epsfbox{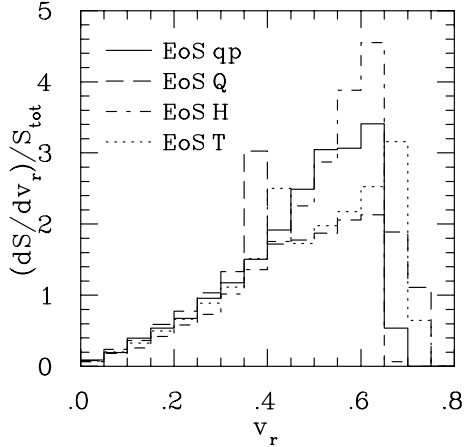}
  \end{center}
  \caption{Entropy flow through fluid elements on decoupling
           surface as function of the transverse flow velocity of each
           fluid element.}
  \label{entropy}
\end{figure}

Even if the flow velocity distributions shown in Fig.~\ref{profiles}
do not look too different from each other, the amount of particles
emitted from fluid elements at different velocities is very different.
To characterise this, the velocity curves in Fig.~\ref{profiles} are
divided into segments so that each segment corresponds to 20\% of
entropy flowing through surface and thus $\sim 20$\% of particles
emitted.  Also the entropy flow as function of flow velocity on the
decoupling surface is shown in Fig.~\ref{entropy}. As can be seen
EoS\,Q leads to very different distribution with much more particles
being emitted at small flow velocities. Especially the ``shoulder'' in
flow profile around $r = 5$ fm leads to a peak in entropy distribution
at $v_r \approx 0.38$ whereas EoSs qp and H lead to distributions
peaking at $v_r\approx 0.6$, close to maximum values of flow
velocity. EoS\,T is again a compromise between these two extremes. The
entropy flow has a peak both at the ``shoulder'' at $v_r\approx 0.42$
and close to maximum velocity at $v_r\approx 0.68$. The largest flow
velocity is also larger than for EoSs qp and H and close to the
maximum for EoS\,Q. The flow on decoupling surface is thus weighted
very differently for each EoS and the average values of flow velocity
and anisotropy do not completely describe $p_T$ differential
anisotropies of particles.

In Ref.~\cite{Derek-RQMD} similar velocity profiles were considered
linear and corroborating the general use of linear velocity profiles
in hydrodynamically inspired fits to particle spectra. As seen here
the deviations from linear behaviour are important at least in
non-central collisions. Thus the parameter values from fits can
deviate from values obtained in full-fledged hydrodynamical
calculations.

\section{Conclusions}

In this paper we have examined how the order of deconfinement phase
transition affects the anisotropy in a hydrodynamical description of
relativistic nuclear collision. We used four different Equations of
State -- one lattice inspired EoS with a crossover transition from
hadronic to partonic phase (EoS\,qp), one where a simple Maxwell
construction between different phases creates a first order phase
transition (EoS\,Q), a purely hadronic EoS with no phase transition at
all (EoS\,H) and an EoS where different phases were smoothly connected
with a hyperbolic tangent function (EoS\,T).

The $p_T$ distributions of various particles could be reproduced
equally well using each of these EoSs when the freeze-out density was
chosen accordingly. Our result is thus different from
Ref.~\cite{Derek-RQMD} where $p_T$ distributions were sensitive to the
amount of latent heat of the EoS. This difference is due to different
treatment of freeze-out. In Ref.~\cite{Derek-RQMD}, the hadronic stage
was described using RQMD cascade model which does not have freeze-out
temperature or density as a free parameter.

The main sensitivity to the EoS was seen in the differential
anisotropy of heavy particles at low $p_T$, i.e. antiprotons and
lambdas.  None of the EoSs was able to reproduce the data, but the EoS
with the first order phase transition, EoS\,Q was closest. Surprisingly
the lattice based EoS\,qp was as far from the proton $v_2$ data as the
EoS\,H without any phase transition. The basic rule was that the
sharper the rapid rise in entropy and energy density at phase
transition and the larger the latent heat, the lower the differential
anisotropy of antiprotons at low $p_T$ was. This, however, is valid
only among the EoSs discussed here.

The results here favour EoS\,Q and first order phase transition over
lattice inspired EoS\,qp. One should not interpret this to mean that
hydrodynamical description of elliptic flow requires a first order
phase transition since EoS\,T with a crossover transition lead to only
marginally worse results than EoS\,Q. The main difference between EoSs
qp and T is in the size of the increase in energy and entropy
densities around the critical temperature and consequently how wide is
the region where the speed of sound is small. Thus the acceptable
description of elliptic flow seems to require very fast and
sufficiently large increase in entropy and energy densities around
$T_c$.

However, these results must be taken as only preliminary. For
simplicity hadron gas was assumed to maintain chemical equilibrium
until kinetic freeze-out in these calculations. As mentioned in
section~\ref{init}, this assumption does not allow the reproduction of
observed particle yields but only the slopes of their spectra and
approximatively their anisotropies~\cite{Heinz,letter1}. The recent
calculations where this assumption is relaxed and a separate chemical
and kinetic freeze-outs included in the model~\cite{Hirano,Peter-Ralf},
have lead to much worse description of the data~\cite{phenixwhite}.
It looks like it is very difficult to describe the data using ideal
fluid hydrodynamics while the hadron gas is not in chemical
equilibrium~\cite{miklos-tetsu}. On the other hand, if the hadronic
phase is described using RQMD cascade which allows chemical
non-equilibrium, the data is again reproduced~\cite{Derek-RQMD}. Thus
the correct treatment of the hadronic phase in a hydrodynamical model
is an open question and it is not yet possible to draw final
conclusions about the details of the EoS based on the observed
anisotropies.


Nevertheless our results point to that a large and rapid increase in
densities around critical temperature is necessary in hydrodynamical
description to describe the observed anisotropies. The failure of
lattice inspired EoS to do this raises the questions whether the
lattice result used here is sufficiently accurate around $T_c$,
whether the hadron resonance gas description of the EoS below $T_c$ is
inaccurate or whether some finite size effects make the EoS relevant
for heavy ion collisions differ from lattice QCD results.

\section*{Acknowledgements}
I am grateful for Thorsten Renk and Roland Schneider for allowing us
to use a parametrisation of their quasiparticle EoS. Fruitful
discussions with P.~V.~Ruus\-kanen, R.~Snellings and S.~S.~R\"as\"anen
are also thankfully acknowledged. This work was partially supported by
the Academy of Finland under project no.\ 77744.

\end{document}